\documentclass[10pt, aps, prl, superscriptaddress, nofootinbib, 
               amsmath, 
               twocolumn, preprintnumbers, 
               showpacs, 
               raggedbottom, 
               floatfix, 
               longbibliography, 
               colorlinks=true, linkcolor=blue, citecolor=blue, urlcolor=blue, 
              footinbib
              ]{revtex4-1}
\usepackage[T1]{fontenc}
\usepackage[utf8]{inputenc}
\usepackage[secnoindent, 
            ]{my_paper} 
\usepackage{my_acronyms}        
\usepackage[english]{babel}     
\usepackage{adjustbox}          
\usepackage{colortbl}
\usepackage{xparse}             
\usepackage[caption=false]{subfig}
\usepackage[percent]{overpic} 

\usepackage[
            tightpage]{preview}

\renewcommand{\vect}[1]{\boldsymbol{#1}}

\newcommand{\beq}{\begin{equation}}
\newcommand{\eeq}{\end{equation}}
\newcommand{\bea}{\begin{eqnarray}}
\newcommand{\eea}{\end{eqnarray}}

\begin{document}

\newlength{\mybaselineskip}
\setlength{\mybaselineskip}{\baselineskip}

\title{Unitary evolution with fluctuations and dissipation}

\author{Aurel Bulgac}%
\affiliation{Department of Physics,  University of Washington, Seattle, Washington 98195--1560, USA}

\author{Shi Jin}%
\affiliation{Department of Physics,  University of Washington, Seattle,  Washington 98195--1560, USA}

\author{Ionel Stetcu}%
\affiliation{Theoretical Division, Los Alamos National Laboratory, Los Alamos, NM 87545, USA}

\date{\today}
\preprint{NT@UW-18-04, LA-UR-18-24160}
 
\begin{abstract}

We outline an extension of the classical Langevin equation  to 
a quantum formulation of the treatment of dissipation and fluctuations 
of all collective degrees of freedom with unitary evolution of a many-fermion 
system within an extension of the time-dependent density functional theory. We illustrate the
method by computing the distribution of fission fragment yields for $^{258}$Fm in a quantum hydrodynamic approach and 
a typical trajectory with full unrestricted density functional theory augmented with dissipation and fluctuations.
  
\end{abstract}

\glsreset{NEDF}
\glsreset{GDR}
\glsreset{DFT}
\glsreset{UFG}

\maketitle


\glsreset{NEDF}
\glsreset{GDR}
\glsreset{DFT}
\glsreset{UFG}

{\bf \emph{Introduction}}\\

The description of the dynamics of a small system in interaction with
a very large reservoir is one of the oldest problems in many-body
physics, starting perhaps with the 1828 work of Robert Brown on
Brownian motion, followed by the illustrious theoretical studies of
Einstein, Smoluchowski, Langevin, Fokker, Planck, Kramers, and a great number of
many others~\cite{Kampen:1992}.  If there are no memory effects, for a
Brownian particle in the Markov approximation one can use the
stochastic Langevin equation, which in one dimension reads:
\beq
m\ddot{x}(t)= F(x(t)) -\gamma m \dot{x}(t) +m\xi(t).  \label{eq:langevin}
\eeq
Here $\xi(t)$
is a zero mean (real) Gaussian white noise with variance
$ \langle\!\langle \xi(t)\xi(t')\rangle\!\rangle = \Gamma
\delta(t-t')$
and angle brackets stand for statistical averaging.  The strength of
the damping and of stochastic forces are related by the Einstein
dissipation fluctuation theorem, $ m\Gamma =2\gamma\text{T}$, where
$\textrm{T}$ is the temperature.  (The Boltzmann constant is chosen $k_B=1$.) 
Appropriate implementations of the
Langevin equation or the equivalent Fokker-Planck equation have been used in nuclear physics for decades, see
Refs.~\cite{Grange:1983,Abe:1986,Frobrich:1998,Moller:2001,Ishizuka:2017,Sierk:2017,Sadhukhan:2017}
and earlier references therein, in order to describe the dissipative
character of the heavy-ion collisions at intermediate energies, the
fission fragments yields, etc.  Recently the Smoluchowski equation has been suggested as a 
simpler alternative to the Langevin approach by Randrup 
et al.~\cite{Randrup:2011,Randrup:2011a,Randrup:2013,Ward:2017,Randrup:2018}, 
which assumes that the collective dynamics is overdamped, in complete agreement with 
our recent findings in a fully microscopic treatment of the fission dynamics at the mean field level~\cite{Bulgac:2018}.

In the case of a quantum mechanical
system one aims to replace the Schr\"odinger equation with a master
equation~\cite{Haake:1973}
\beq
i\hbar \dot{\rho} = [H,\rho]+{\cal L}(\rho), \label{eq:BGKKY}
\eeq
where $\rho=\text{Tr}_\text{res}\; \rho_\textrm{tot}$ is the density matrix of
the subsystem after taking the trace over the reservoir/intrinsic coordinates of
the full $\rho_\textrm{tot}$, $H$ is the Hamiltonian of the collective/isolated
system, and ${\cal L}(\rho)$ is a linear super-operator acting on
$\rho$.  Such an equation for the one-body density matrix can be formally 
derived from a BGKKY hierarchy~\cite{Huang:1987}.
The form of the super-operator is difficult to use in practice, it is
generally non-local in time, and the emerging master equation is very
difficult to solve.  The dynamics of the system is entangled with the
dynamics of the reservoir. However, in a Markov approximation one 
can derive a generalization of the von Neumann~\cite{Neumann:1927}
and Landau equation~\cite{Landau:1927}.
The most
general form of the master equation in a Hilbert space of dimension
$N$ is much simpler than Eq. \eqref{eq:BGKKY} and of the form~\cite{Gorini:1976,Lindblad:1976},
and it is routinely referred to as the Lindblad equation,
\bea
& i\hbar \dot{\rho}= 
[H,\rho] -i[W\rho+\rho W]+i\sum_{k,l=1}^{N^2-1} h_{kl} A_k \rho A^\dagger _l, \label{eq:lind}\\
&W =  W^\dagger= \frac{1}{2}\sum_{k,l=1}^{N^2-1} h_{kl} A^\dagger _l
A_k, \label{eq:lind2}
\eea
where $h_{kl}$ and $W$ are Hermitian positively defined matrix and
operator respectively, and $A_k$ form a full set of linearly
independent operators, apart from the unit operator. This equation was derived by 
requiring the preservation of  the total probability ($\text{Tr}\dot{\rho}\equiv 0$) and
of the positivity of $\rho$ during the time evolution.  Simpler and/or equivalent master equations have
been derived over the years in quantum optics in perturbation
theory~\cite{Bloch:1946,Mollow:1975,Gisin:1984,Collett:1984,Dalibard:1992,Dum:1992,Molmer:1993,Carmichael:1993,Daley:2014}.
If one were to drop the last term in Eq.~\eqref{eq:lind} the
probability would not be conserved, as
$\hbar \,\text{Tr}\dot{\rho} = -2\text{Tr}(W\rho)\le 0$.  $W$ plays the
role of an optical potential, thus being responsible for simulating
dissipation, which has been used in either one-channel or coupled
channels situations.  The loss of probability is only marginally alleviated
in coupled channel treatments (where optical potentials are also present), 
when the probability from the incoming
channel is only partially recovered in the other channels.  

The direct numerical solution of the Lindblad equation \eqref{eq:lind} or of its Monte
Carlo wave function
formulation~\cite{Gisin:1984,Collett:1984,Gardiner:1985,Gardiner:1986,Gisin:1992,Gisin:1993,
  Dalibard:1992,Dum:1992,Molmer:1993,Carmichael:1993,Daley:2014} can turn into a
formidable problem in cases of interest in nuclear physics, where the
number of degrees of freedom (DoF) in Langevin studies is between 2 to
at most
5~\cite{Frobrich:1998,Moller:2001,Randrup:2011,Ishizuka:2017,Sierk:2017,Sadhukhan:2017},
which would correspond in the case of a quantum treatment to wave
functions of 2 to 5 variables and density matrices depending on 4 to
10 spatial variables alone, plus time, which easily becomes prohibitive numerically.  
We should mention here that over the years many
extensions of the time-dependent meanfield approaches have been
suggested in nuclear physics, in order to incorporate fluctuations in
a quantum treatment~\cite{Reinhard:1992,
  Reinhard:1992a,Juillet:2004,
  Lacroix:2005,Lacroix:2007,Lacroix:2008,Ayik:2008,Tanimura:2017}.
  
 In this work, instead of limiting the number of collective DoF to a
small number of chosen characteristics or moments of the number
density $n(\bm{r},t)$, we will consider the entire number density as
our chosen set of collective DoF and treat them in a  quantum
formalism at a ``finite temperature,'' which is controlled by the
intrinsic DoF.  We argue that one can add carefully chosen additional
terms to the usual TDDFT
equations to simulate both dissipation and fluctuations of the nuclear
collective motion and maintain at the same time the unitary character
of the evolution, a distinctive characteristic of TDDFT. Since fluctuations are random,
observables will have to be evaluated as ensemble averages over these
realizations. However, in the case of steady-state situations one can also consider 
time-averaged observables, which should lead to the same final results.\\
  
  {\bf \emph{Difficulties with  introducing dissipation and fluctuations in meanfield dynamics}}   \\

Numerically,
Lindblad equation is usually solved by means of a Monte Carlo wave function,
obtained with a Hamiltonian $H-iW$ augmented with a stochastic term,
which simulates the role of the last term $\sum_{k,l=1}^{N^2-1} h_{kl} A_k \rho A^\dagger_l$ in
Eq.~\eqref{eq:lind}~\cite{Gisin:1984,Collett:1984,Gardiner:1985,Gardiner:1986,Gisin:1992,Gisin:1993,
  Dalibard:1992,Dum:1992,Molmer:1993,Carmichael:1993}.  
 The corresponding augmented stochastic ``Schrödinger'' equation 
 for the single quasi-particle wave functions in a 
 time-dependent density functional theory (TDDFT) 
 would acquire then the form
 \beq
 i\hbar\dot{\psi}_k= [H-iW]\psi_k +S\psi_k+\sum_l\lambda_{kl}\psi_l
\eeq
where $S$ is in general a non-hermitian complex stochastic field, appropriately defined
and $\lambda_{kl}$ are Langrange multipliers enforcing 
at all times the orthogonality conditions $\langle\psi_k|\psi_l\rangle = \delta_{kl}$.
Unfortunately this approach cannot be used in simulating fermion systems, where one needs 
to evolve in time a large number of single quasi-particle wave functions $\psi_k$. 
If these orthogonality conditions are satisfied only after ensemble averaging instead 
one would introduce large unphysical
particle number fluctuations~\cite{Bulgac:2018}, apart from expected collective energy fluctuations. 
With particle number fluctuations present one would have a difficult task in quantifying their role in the definition 
of the width of the energy fluctuations. 

Upon considering dissipation and fluctuations one expects fluctuations 
of the collective energy, but the total particle number should be exactly conserved.
The presence of the ``optical potential'' $-iW$ (and of a non-hermitian  stochastic field $S$) leads to a 
non-unitary evolution, which could in principle be restored only by introducing a large 
number of time-dependent Lagrange multipliers $\lambda_{kl}$, rendering this system of 
such coupled equation basically impossible to handle numerically.

The unravelling of the Lindblad equation~\eqref{eq:lind} 
 to a stochastic or Monte Carlo form is not a unique procedure and 
 one may introduce various other forms for the ``optical potential $-iW$,'' in which case 
 $W$ does not have to be hermitian for example. One can also try to work 
 with non-orthonormal quasi-particle wave functions, at the 
 expense of making the evaluation of observables 
 much more challenging numerically in the case of fermionic systems. 
 Allowing for total particle fluctuations can lead to spurious mass and 
charge distributions in fission dynamics for example. 

Our goal here is to construct a quantum approach  equivalent to
the classical Langevin approach. 
In nuclear applications of the Langevin approach
practitioners typically select a few characteristics of the
number density $n(\bm{r},t)$ (e.g. in the case of fission: elongation
of the nucleus, mass asymmetry, neck size, and two quadrupole
deformations of the fragments~\cite{Moller:2001}).  For these
collective variables one constructs a potential energy surface by minimizing 
the total energy with suitable constraints, an
inertia and a dissipation tensor and assume the existence of coupled
Langevin equations for these collective variables. The potential energy 
and the inertia tensor are 
constructed strictly speaking at zero temperature and the evolution 
of the intrinsic DoF is  assumed to be adiabatic, thus with no 
intrinsic excitations, at local zero temperature when the collective variables are kept fixed. 

The adiabaticity
assumption in large amplitude collective motion in nuclear physics, typically conflated with 
the slowness of the collective motion,
translates into no entropy production for the intrinsic variables, $S_\text{int}(t)\equiv 0$.  
Since dissipation is included only in the collective motion DoF 
their entropy naturally increases in a Langevin approach, $\dot{S}_\text{coll}(t)>0$.
At the same time, the total entropy of the nuclear system, which is isolated,  
should in principle exactly vanish at all times, $S_\text{tot}(t)\equiv0$. 
This apparent contradiction finds its resolution, not in a Langevin approach to collective motion,
but in an approach which allows continuous energy exchange from the collective 
to the intrinsic DoF, which on relatively long time scales of interest in nuclear dynamics appears irreversible. 
In that case one can establish that the entanglement entropy of the intrinsic DoF,
$S_\text{int}(t)=-\text{Tr}_\text{coll} \left [\rho(t) \ln \rho(t)\right ]$, 
actually increases $\dot{S}_\text{int}(t)>0$. 
In a TDDFT approach to nuclear fission, 
even in the absence  of explicit dissipation in the collective DoF the relation $\dot{S}_\text{int}(t)>0$
is automatically fulfilled, as energy is practically 
irreversibly transferred from collective to intrinsic DoF~\cite{Bulgac:2018}. 
Entropy is strictly speaking a quantity which is obtained only after 
the system  ``explored the relevant" part of the phase-space
and the time average became equal to the corresponding phase-space average~\cite{LL5:1980}.
 During the descend from saddle-to-fission a nucleus 
might not necessarily have enough time to relax and the 
intrinsic entropy might have not reached its equilibrium value at each fixed values 
of the collective coordinates.

The number and character of the collective
variables are chosen according to various authors preferences 
or arguments, which are not uniformly endorsed.
There exist rather compelling theoretical arguments that the actual
number of relevant DoF is (much) greater than considered so far. In fission 
for example, on the top of the barrier the intrinsic excitation energy 
is rather small and the number of possible excited DoF is also small, but 
that is not true anymore by the time the nucleus reaches the scission 
configuration~\cite{Bulgac:2018}. This
aspect however has not been satisfactorily  settled in literature.  

The classical
Langevin and quantum Lindblad equations both assume the Markov
approximation, but at the same time they differ in one critical
aspect. In the Langevin equation there are two parameters: one,
$\gamma$, which controls the strength of the dissipative force, and a
second one, $\Gamma$, which controls the strengths of the
fluctuations. While their ratio is controlled by the temperature 
$\text{T}$, the absolute value
of each of these parameters control the rate of energy exchange. 
In Lindblad equation \eqref{eq:lind} however the ration of the strengths of the ``dissipation''
$W= \tfrac{1}{2}\sum_{k,l=1}^{N^2-1} h_{kl} A^\dagger _l A_k$ and of
the ``fluctuations'' $\sum_{k,l=1}^{N^2-1} h_{kl} A_k \rho A^\dagger_l$
is fixed and independent of the temperature.
\textcite{Diosi:1993} and \textcite{Akamatsu:2015,Akamatsu:2015a,Kajimoto:2018,Akamatsu:2018} have
demonstrated however how in the high temperature limit one can
introduce independently dissipation and fluctuations in the Lindblad
equation.  Efforts are under way to generalize this type of
approach to finite
temperatures~\cite{Diosi:1998,Strunz:1999,Strunz:1999a,Strunz:1996,Diosi:1997,
  Diosi:1998,Strunz:1999,Strunz:1999a,Strunz:2005,Strunz:2005a},
following the Feynman-Vernon~\cite{Feynman:1963} and
Caldeira-Leggett~\cite{Caldeira:1983} formalisms.

 While Lindblad equation appears as the most natural extension of a classical Markovian 
 evolution to a quantum one, one aspect has not been resolved in the literature as far as we know.
 When coupling a classical system to a reservoir one expects that after a certain relaxation time the system arrives 
 at an equilibrium. In the case of a quantum system one would expect the time averaged density matrix 
 to reach asymptotically the limit $\rho \propto \exp\left ( -\frac{H}{T}\right )$. In the case of the Lindblad equation \eqref{eq:lind}
 that would be equivalent to the condition that
 \beq
 \sum_{k,l=1}^{N^2-1} h_{kl} \left [ A^\dagger _l    A_k H+H  A^\dagger _l   A_k   \right ]
 \equiv 2\sum_{k,l=1}^{N^2-1} h_{kl} A_k H A^\dagger _l,
  \eeq
 which would imposed rather serious constrains on the set of operators $A_k$. 
It is not clear under what conditions a steady-state solution $\rho \propto \exp\left ( -\frac{H}{T}\right )$
can be achieved, unlike in the case of Fokker-Planck, Langevin or Boltzmann equations. It is not obvious
 what would be in general the steady-state solution of the Lindblad equation. Unlike in 
 the case of the Langevin equation \eqref{eq:langevin} there is no general prescription on how one might control
the equilibrium temperature of the quantum system.

While Lindblad approach has many appealing mathematical features, attempting to use it for  
describing a nuclear system appears to lead in the best case scenario to a very cumbersome formalism.  
We propose to go in a different direction: i) first to remove the somewhat artificial limitation of 
phenomenological models to a small and somewhat arbitrary number of collective variables;
ii) second to develop a formalism in which the introduction of dissipation leads to a unitary 
evolution of the single-particle DoF and time evolution of the single-particle density satisfies 
the continuity equation; iii) the energy increase due to ``random'' fluctuations
is properly balanced by the dissipation; iv) a formalism in which one can control separately the rate of fluctuations and 
dissipation as well as the temperature of the stationary state (as in the case of the classical 
Einstein fluctuation-dissipation theorem). This particular aspect is relevant in non-equilibrium 
processes, when the equilibration time can be longer than the time at which the system 
changes its ``macroscopic'' properties. 

While the use of Gaussian white noise with either the It\^{o} or 
Stratonovich calculus is mathematically extremely
seductive~\cite{Kampen:1992,Kampen:1981}, and has been advocated over
the years in order to devise various generalizations of the meanfield
dynamics, fluctuations in space and time corresponding to arbitrarily
large fluctuations in momenta and energy are physically unjustified.
Fluctuations in
nuclear collective motion on a spatial scale much smaller than the
average interparticle
distance, which is of the order of 2 fm for nuclei,
or on a temporal scale shorter than a few 10's fm/c (comparable to the
the time it takes the fastest nucleon to traverse a big nucleus) are
unwarranted.  The numerical implementation of stochastic differential
equations has subtleties and is more difficult to carry out than in case of
differential equations. At the same time, in actual numerical
implementations the high frequencies (inherent for white noise) are
eliminated by using finite integration time-steps, which is equivalent
to performing a short time coarse graining of the ``true'' numerical
solution.  Instead of introducing coarse graining in either time or space 
dictated by the numerical implementation 
we will set physics inspired limits on the
character of fluctuations.  Dissipation in collective motion in
low-energy nuclear dynamics is mostly one-body in
character~\cite{Blocki:1978} and very strong at the same time. These aspects 
were firmly confirmed recently in an unrestricted 
implementation of TDDFT
quantum microscopic framework~\cite{Bulgac:2018}, 
without resorting to introducing the hard to 
define both the number and the character of the collective DoF, collective inertia, 
potential energy surface, or friction mechanisms.  
This is  a mechanism similar to
Fermi's model for high-energy cosmic rays~\cite{Fermi:1949}.  In
low-energy induced nuclear fission the two-body dissipation mechanism
is inhibited due to the relatively small phase space available, 
corresponding to excitation energies less than about 20 MeV, and the 
corresponding long nucleon mean free path~\cite{Bohr:1969}.

The time-dependent meanfield equations can be formally obtained within
a path integral approach of the propagator, as described by
\textcite{Negele:1988}, using the stationary phase approximation of a
path integral representation of the many-body propagator
$\int {\cal D}\sigma\exp\left (iS[\sigma]/\hbar\right )$, where
$S[\sigma]$ is the action.  There is a consensus that the TDDFT
description provides a description of the average or more likely of the most probable
dynamics, thus the same kind of trajectory obtained in the stationary
phase approximation of the path integral~\cite{Bulgac:2010}.  What is missing in TDDFT is
the contribution from fluctuations, which formally would appear as
an additional contribution $g_2$ to the equation for the one-body
density matrix $\rho_1$:
\beq
i\hbar\dot{\rho}_1 - [h_1(\rho_1),\rho_1] = g_2.
 \label{eq:bgkky}
 \eeq
In the meanfield approximation this type of equation 
can be obtained from the  BGKKY hierarchy~\cite{Huang:1987}. 
 In the limit when $g_2$ vanishes this reduces to the time-dependent
 meanfield approximation for the one-body density matrix $\rho_1$.
 The semiclassical limit of Eq.~\eqref{eq:bgkky} reduces
 to the Boltzmann equation, if $g_2$ is approximated with the
 collision integral.  Upon taking a short time average the term $g_2$ vanishes.  
 It would appear natural to generalize
 Eq.~\eqref{eq:bgkky} to some kind of stochastic form, to incorporate
 to the role of the neglected, and relatively rapid fluctuations of
 $g_2$.  
 
 The TDDFT equations for the single-particle wave functions
 are obtained using a nuclear energy density functional (NEDF).  The
 NEDF should satisfy the local Galilean covariance, which implies that
 the total energy of the system can be represented as a sum~\cite{Engel:1975,Bender:2003,Bulgac:2013a}
\bea
&& E_\text{tot}(t) \!\!=
E_\text{coll}(t)+E_\text{int}(t)
\equiv \int \!\!d^3 {\bf r}\frac{ mn({\bf r},t){\bf v}^2({\bf r},t)}{2} \nonumber \\
&& +\int \!\!d^3{\bf r}\, {\cal E}\left (\tau({\bf r},t)-n({\bf r},t)m^2 {\bf v}^2({\bf r},t), n({\bf r},t),...\right ),\label{eq:etot}
 \eea
 where $n({\bf r},t)$ is the number, $\tau({\bf r},t)$ is the kinetic, and 
 ${\bf p}({\bf r},t)=mn({\bf r},t){\bf v}({\bf r},t)$ are linear 
 momentum and local collective/hydrodynamic velocity
 densities, and ellipses stand for various other densities.  The first
 term in Eq.~\eqref{eq:etot} is the collective/hydrodynamic energy
 flow $E_\text{coll}$ and the second term is the intrinsic energy
 $E_\text{int}$ in the local rest frame.  For the sake of simplicity
 we have suppressed the spin and isospin DoF, even though they are included
 in the numerical examples discussed below. \\

{\bf \emph{Augmented nuclear TDDFT equations including dissipation and fluctuations} }\\

 The TDDFT evolution equations augmented to incorporate dissipation
 and fluctuations we introduce have the form
\bea \label{eq:fd}
&&i\hbar \dot{\psi}_k({\bf r},t)= h[n]\psi_k({\bf r},t) + \gamma[n] \dot{n}({\bf r},t)\psi_k ({\bf r},t)\\
&& -\frac{1}{2} \left [ {\bf u}({\bf r},t)\cdot \hat{{\bf p}}+ \hat{{\bf p}} \cdot {\bf u}({\bf r},t) \right ] \psi_k({\bf r},t)
+ u_0 ({\bf r},t) \psi_k ({\bf r},t), \nonumber
\eea 
where $\hat{\bm{p}}=-i\hbar\bm{\nabla}$ (not to be confused with the linear momentum density 
${\bf p}({\bf r},t)$), the index $k$ runs over the neutron and proton
quasi-particle states and where $\psi_k({\bf r},t)$ are 4-component 
quasi-particle wave functions and $h[n]$ is a $4\times4$ partial 
differential operator~\cite{Bulgac:2016,Bulgac:2018}.  The fields ${\bf u}({\bf r},t)$ and $u_0 ({\bf r},t) $ 
generate both rotational and irrotational dynamics. 
One can also introduce a carefully
  chosen fluctuating inertia tensor
  ${\bf p}\cdot\tensor{\text T} ({\bf r},t) \cdot {\bf p}$, a fluctuating spin-orbit 
  interaction, a fluctuating pairing field $\delta({\bf r},t)$,  
  and a time-symmetry 
  breaking stochastic field ${\bm \sigma} \cdot {\bf C}({\bf r},t)$. Basically, 
  every term of the quasi-particle Hamiltonian can be rendered stochastic.
  
The term
$\gamma[n]\dot{n}\propto - {\bm \nabla}\cdot {\bm p}({\bm r},t)$, per
continuity equation~\cite{Bulgac:2013c}, plays the role of ``quantum
friction'', in which the ``friction coefficient'' $\gamma[n]$ can
depend on the number density and/or its gradient, etc. and thus it can
simulate volume and/or surface "friction."  In the presence of this
``quantum friction'' term alone $\dot{E}_\textrm{tot}(t)\le 0$ and
$\lim_{t\rightarrow \infty} {\bf v}({\bf r},t)=
0$~\cite{Bulgac:2013c}, similarly to the classical Langevin equation.  
We should add that over the years many authors have discussed 
various other forms of ``quantum friction'' extensions of the 
Schrödinger equation~\cite{Immele:1975,Kostin:1972,Hasse:1975,
Kanai:1948,Caldirola:1941,Albrecht:1975,Tokieda:2017}, some of which have 
similarities with our suggested form for the ``friction' potential.'' However, these 
earlier suggestions introduce typically averages of either momenta or coordinates 
over the entire system, thus introducing unphysical non-localities into the theory. 
It makes no sense to have the magnitude of the dissipation in one part of the system 
depend on the properties of another  part of the system, which can be 
spatially separated by a large distance.

The field
$u_0 ({\bf r},t)$ and each of the Cartesian components of the 3D
velocity field ${\bf u}({\bf r},t)$ are uncorrelated stochastic fields
of the type to be described below, see Eq. \eqref{eq:chi}.  By construction
Eqs. \eqref{eq:fd} lead to a unitary evolution with dissipation built
in (unlike the case of an optical potential). The "quantum friction"
term and the stochastic fields do not affect the relative momenta of
any pair of nucleons, and therefore they do not contribute to the
thermalization of the intrinsic motion.  The average value of
the total local/collective momentum ${\bf p}({\bf r},t)$ is modified
by the ''quantum friction'' and these additional stochastic fields,
and thus only the collective DoF (moments of the density)
are affected, as in case of the Langevin approach~\cite{Frobrich:1998, 
Moller:2001,Randrup:2011,Ishizuka:2017,Sierk:2017,Sadhukhan:2017}. 
There is a one-to-one correspondence between 
the neutron and proton number densities $n_{q}({\bf r},t)$ with $q=n,p$
and all their possible moments and we call them the collective DoF.
They define the shape of the nucleus, in full analogy with the generator 
coordinate method~\cite{Hill:1953,Griffin:1957} or the extended generator coordinate method~\cite{Goeke:1980}. 
 
The TDDFT dynamics automatically incorporates the one-body dissipation
mechanism~\cite{Blocki:1978}.  The additional ``quantum
friction'' term is needed to counteract the heating due to the
stochastic fields $u_0 ({\bf r},t)$ and $m{\bf u}({\bf r},t)$.  The
strength of the ``quantum friction'' should be chosen in analogy to
the Einstein fluctuation-dissipation theorem.  In the case
of a Brownian particle, one follows the dynamics of the Brownian
particle alone, but not the effects on the dynamics of the fluid  
and the total energy of the
fluid and Brownian particle are not conserved.
In a TDDFT augmented with dissipation and fluctuations one follows
the coupled dynamics of both collective and intrinsic DoF
within a stochastic framework. 
  
The generic time-dependent 3D field structure of both the scalar $u_0({\bf r},t)$ and
of each cartesian component of the vector stochastic fields ${\bf u}({\bf r},t)$ is of
the form for $\nu= 0, x,y,z$:
\bea
& u_\nu ({\bf r},t)= \sqrt{\Gamma}  \sum_{k=1}^{N_k} F(t-t_k,\tau_k)  \eta_k({\bf r}), \label{eq:chi} \\
&\eta_k({\bf r})= \sqrt{\frac{1}{N_{kb}}} \sum_{l=1}^{N_{kb} } \alpha_{kl} G({\bf r}-{\bf r}_{kl}, a_{kl}), 
\eea
where $F^2(t,\tau)$ and $G^2({\bf r},a)$ are 1D and 3D 
smoothed normalized $\delta$-functions of width $\tau$ and $a$ respectively. The finite widths $\tau$ and $a$
impart these stochastic fields a finite memory time and a finite correlation length respectively. Here
$\langle t_k-t_{k-1} \rangle \propto  \langle \tau_k\rangle = {\cal O}\left (\frac{ mr_0A^{1/3}}{\hbar k_F}\right )$,  
$\langle N_{kb} \rangle = {\cal O}(A)$,
$ \langle \alpha_{kl}\rangle =0$, 
$\langle\! \langle \alpha_{kl}\alpha_{mn}\rangle\!\rangle  = \delta_{km}\delta_{ln}$, 
$\langle a_{kl} \rangle = {\cal{O}}\left ( \frac{\pi}{k_F}\right )$, $\langle |{\bf r}_{kl}|\rangle  ={\cal O}(r_0A^{1/3})$, 
$r_0\approx 1.2$ fm, $A$ is the mass number, $k_F$ is the Fermi
momentum, and $\Gamma$ is a parameter controlling the variance of the
$u_\nu({\bf r},t)$, see Eq.~\eqref{eq:G}.  $t_k$, $\tau_k$, $N_{kb}$,
$a_{kl}$, and ${\bf r}_{kl}$ are uncorrelated uniform random numbers
in properly chosen intervals. Then
\bea
&&\int \!\!\! d^3 {\bf r}\langle \eta_k({\bf r})\rangle =0, \; \int \!\!\! d^3{\bf r}\langle \eta_k({\bf r})\eta_l({\bf r})\rangle =\delta_{kl}, \label{eq:k}\\
&&\int_0^t \!\!\!\!\!\! dt' \!\!\!  \int  \!\!\! d^3{\bf r} \langle u_\nu({\bf r},t') \rangle =0,
      \int_0^t \!\!\!\!\!\! dt' \!\!\!  \int  \!\!\! d^3{\bf r} \langle u_\nu^2({\bf r},t') \rangle \approx \Gamma \langle N_k\rangle, \label{eq:G}\\
&&\int_0^t \!\!\!\!\!\! dt' \!\!\!\int \!\!\! d^3{\bf r}\langle  u_\nu({\bf r},t') u_\nu({\bf r},t'+\Delta t) \rangle \approx 0, \; \Delta t \gg \langle \tau_k\rangle ,
\eea
where $t\approx\langle N_k\rangle \langle\tau_k\rangle$.  In the limits
$ \lim_{\tau\rightarrow 0} F^2(t,\tau) = \delta(t)$,
$\lim_{a\rightarrow 0}G^2({\bf r},a)=\delta({\bf r})$,
$\tau_{k}\rightarrow 0$, and $N_{kb}=1$ one recovers the Gaussian
white noise used in typical treatment of stochastic
equations~\cite{Kampen:1981,Kampen:1992}. $u_\nu({\bf r},t)$
simulates random $N_k$ collective ''jolts,'' administered to the
system at random times $t_k$ and of random duration $\tau_k$.  Each
''jolt'' consists of random $N_{kb}$ "bumps/sumps," randomly
distributed throughout the nucleus, each with a
height/depth of zero mean and unit variance and of a random diameter
$a_{kl}$.
The ratio $\Gamma/\gamma\propto \textrm{T}$ controls the temperature
of the intrinsic system, similarly to Einstein fluctuation-dissipation
theorem. 
There are at least two independent coupling strengths
$\Gamma$ (of appropriate dimension), one for ${\bf u}({\bf r},t)$ and
the other for ${u_0}({\bf r},t)$.  

\begin{figure}
\includegraphics[width=0.99\columnwidth]{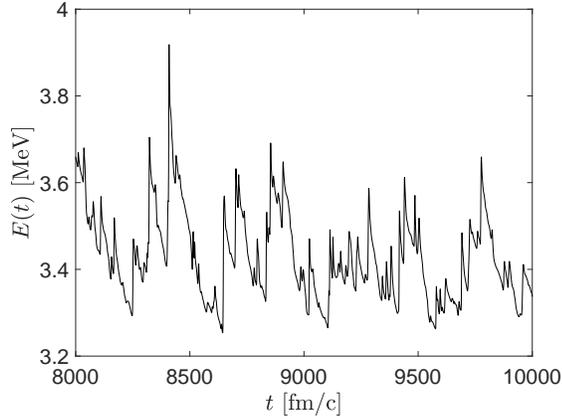}
\caption{ \label{fig:E} The expectation value of the energy of the 1D harmonic 
oscillator as a function of time.   }
\end{figure}

\begin{figure}
\includegraphics[width=0.99\columnwidth]{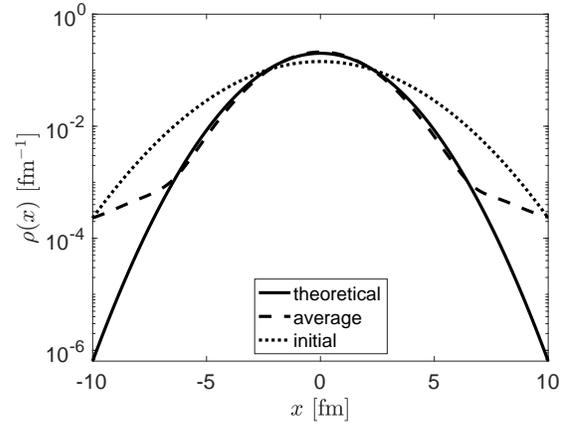}
\caption{ \label{fig:rho} The   initial, final  $\rho(x)= \tfrac{1}{\tau}\int_0^\tau dt |\psi(x,t)|^2$, 
and expected  theoretical density distribution, 
using the temperature estimated from Eq. \eqref{eq:T}.  By increasing the simulation time one can improve 
on the tails of the calculated density distribution.  }
\end{figure}

We illustrate this approach with the case of a nucleon in a 1D harmonic oscillator
$V(x)=\tfrac{m\omega^2x^2}{2}$ with $\hbar\omega = 6$ MeV, a ``quantum friction potential'' as described above and 
and a stochastic field $u_0(x,t)$ only. It is not always obvious that the time-average and the phase-space average, 
or in other words, ergodicity is satisfied in simulations, particularly in the case of integrable 
systems~\cite{Bulgac:1990,Kusnezov:1990,Akamatsu:2018}. 
Starting with a somewhat arbitrary initial state, 
after some time the harmonic oscillator reaches a steady-state solution at a 
temperature $T=1/\beta$ determined from the condition
\bea
&& \frac{1}{\tau}\int_0^\tau dt E(t) = \frac{1}{Z(\beta)}
                      \sum_{n=0}^\infty e^{-\beta\varepsilon_n } \varepsilon_n , \label{eq:T}\\
&& Z(\beta) = \sum_{n=0}^\infty e^{ -\beta\varepsilon_n}, \quad \varepsilon =\hbar\omega \left( n+\frac{1}{2}\right ), \nonumber \\
&& \rho(x) =  \lim_{\tau\rightarrow \infty} \frac{1}{\tau}\int_0^\tau dt |\psi(x,t)|^2\approx  
\frac{1}{Z(\beta)}\sum_{n=0}^\infty e^{-\beta\varepsilon_n}|\phi_n(x)|^2, \nonumber
\eea
where $\phi_n(x)$ are the 1D harmonic oscillator eigenfunctions.

In Fig. \eqref{fig:E} we display the expectation value of the energy of the 1D oscillator as a function of the simulation time,
illustrating that the system attained a steady-state regime. In Fig. \eqref{fig:rho} we show the initial, 
expected, and computed equilibrium density distributions. 
We put a minimal effort into the fine-tuning of  the parameters of the ``quantum friction potential,''
the stochastic field, and of the length of the simulation time.  \\

{\bf \emph{Fission of $^{258}$Fm and $^{240}$Pu}}\\

We illustrate this approach by solving the nuclear
quantum hydrodynamic equations for $^{258}$Fm fission.
At zero
temperature within Landau's two-fluid hydrodynamics only the
superfluid components survive and the dynamics reduces to that of a
neutron and a proton interacting miscible classical perfect/ideal fluids~\cite{Lamb:1945,LL6:1966}
for canonically conjugate fields $n_q({\bf r},t)$ and $\phi_q({\bf r},t)$,
where ${\bm \nabla}\phi_q({\bf r},t)=m{\bf v}_q({\bf r},t)$. 
In a quantum hydrodynamic approach 
we use the semiclassical form of the SeaLL1 NEDF~\cite{Bulgac:2015,Shi:2018}  for homogeneous 
nuclear matter, augmented with a Coulomb energy and gradient terms, 
\beq
{\cal E}_\text{int}={\cal E}_\text{kin}(\tau_n,\tau_p)+ {\cal E}_\text{hom}(n_n,n_p)+
{\cal E}_\text{Coul}+\tfrac{\hbar^2}{2m}(C+Dn)({\bm \nabla} n)^2. \nonumber
\eeq 
which reproduces the symmetric nuclear matter energy, the saturation density, 
the symmetry energy, and the Coulomb energy. $C=-2.8622$ fm$^{3}$ and $D=9$ fm$^{6}$ are chosen to
accurately  reproduce the nuclear surface tension
$\sigma=\int dz\left [ {\cal E}_\text{int}\left (\tau(z),n(z),...\right )  -\mu n(z)\right ]\approx$ 1 MeV/fm$^2$, where $\mu=-15.6$ MeV~\cite{Shi:2018} 
is the chemical potential for infinite symmetric nuclear matter and $n=n_n+n_p$, $n_{n,p}(z)$ and $\tau_{n,p}(z)$ are 
the number and kinetic energy density distribution for semi-infinite symmetric nuclear matter.
Using Madelung representation~\cite{Madelung:1926} of the neutron and proton ''wave functions'' 
$\Psi_q({\bf r},t) = \sqrt{n_q({\bf r},t)} \exp[ i\phi_q({\bf r},t)/{\hbar} ]$,
where ${\bf p}_q({\bf r},t) =n_q({\bf r},t) {\bm \nabla}\phi_q({\bf
r},t)=mn_q({\bf r},t) {\bf v}_q({\bf r},t)$, one can recast the quantum hydrodynamic
equations into two coupled effective Schr\"{o}dinger equations (here with 
$ {\bf u}({\bf r},t) \equiv 0$ for simplicity)  with 
isoscalar dissipation and fluctuations, 
\bea
i{\hbar}\dot{\Psi}_q({\bf r},t) &&= -\frac{{\hbar}^2 }{2m} {\bm \nabla}^2 \Psi_q({\bf r},t) 
   + \frac{ \delta {\cal E}_\text{int} }{ \delta n_q({\bf r},t)} \Psi_q({\bf r},t) \label{eq:td} \\
&&+ \gamma[n] \dot{n}({\bf r},t)\Psi_q ({\bf r},t)\ 
+ u_0 ({\bf r},t) \Psi_q ({\bf r},t). \nonumber
\eea 
The hydrodynamic equations do
not include pairing and shell effects and the ground states for
typical nuclei have spherical symmetry.  Even though we illustrate here a unitary quantum 
evolution with dissipation and fluctuations for two components only, the scale of this simulation
is already significantly larger than any other similar simulation reported in literature so far.

\begin{figure}
\includegraphics[width=1\columnwidth]{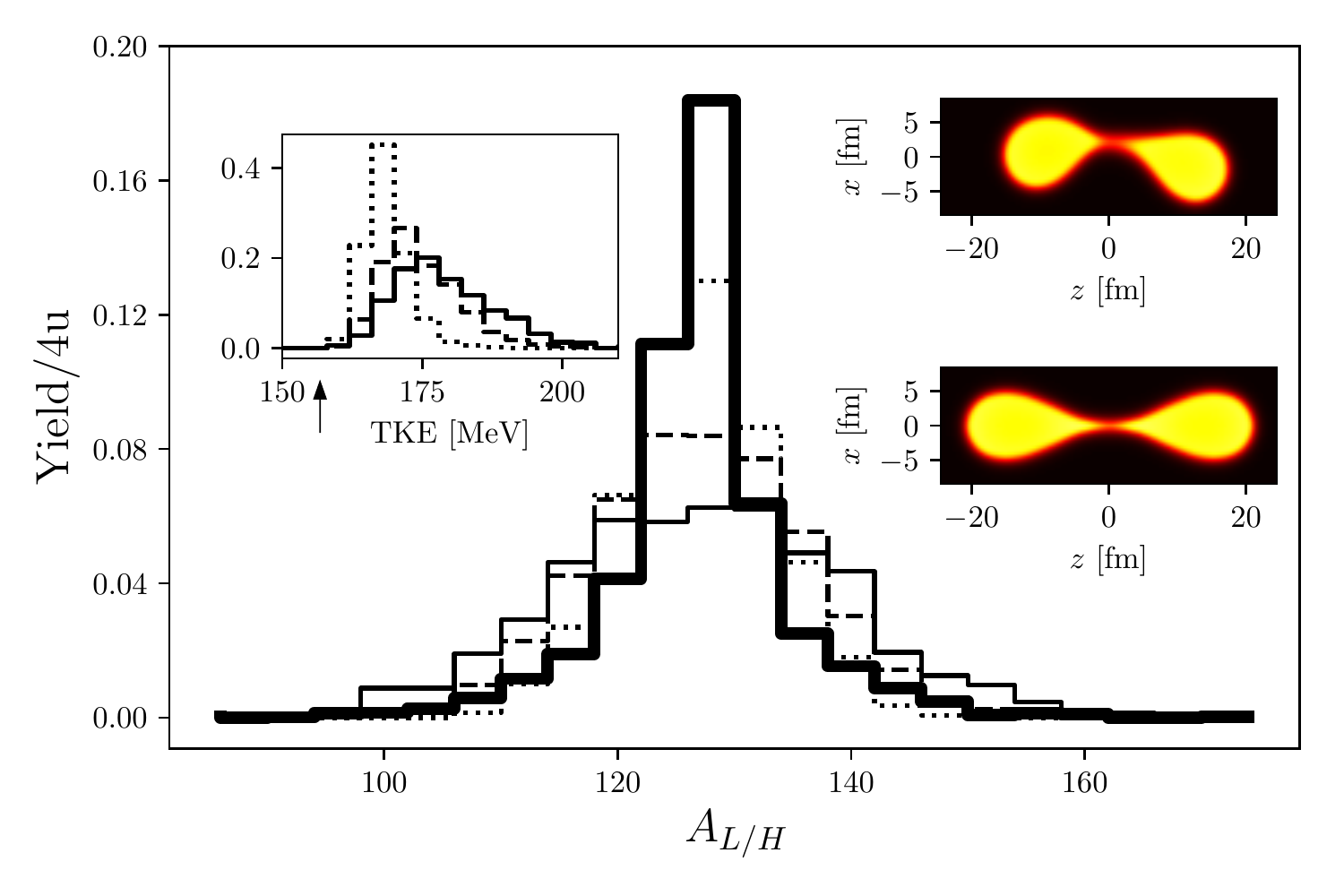}
\caption{ \label{fig:yields} (Color online) The mass yields obtained
  solving the quantum hydrodynamics equations~\cite{Bulgac:2015}
  including dissipation and fluctuations as in Eq.~\eqref{eq:td} at an excitation 
  energy $E^*\approx 14$ MeV, obtained using different strengths 
  $\Gamma$ = 0.1 (solid), 0.05 (dash), and 0.02 (dots) MeV
  of the fluctuating field $u_0({\bf r},t)$ (keeping $\Gamma/\gamma$ fixed),   compared to experimental
  data (thick solid)~\cite{Hulet:1989} for spontaneous fission of $^{258}$Fm. 
  The variance of the mass (24.9, 18.4, and 12.9 amu vs experiment  15.2 amu) and
  of the TKE (18.3, 14.9, and 8.5 MeV vs experiment 19.3 MeV)  
  distributions are approximately proportional to $\sqrt{\Gamma}$. In the insets we show the TKE (left inset)
  distributions  and typical nuclear shapes 
  at scission (right inset) with and without fluctuations and dissipation. 
  The arrow points to the TKE for symmetric splitting  in the absence of 
  dissipation and fluctuations.
  }
\end{figure}

The inclusion of
dissipation and fluctuations easily leads to mass distributions which are
close to the observed ones for spontaneous fission of $^{258}$Fm, see Fig.~\ref{fig:yields}, even
though the goal of this first calculation is only to illustrate the
method.  Even though our calculations are for fission from an excited state 
and the purpose of this figure is to illustrate qualitative aspects of our approach..
The strength of the dissipation and fluctuations terms have
been adjusted to correspond to a ``temperature'' of $\approx 0.75$ MeV, 
or an excitation energy of about 14 MeV.
We have started our simulations with $^{258}$Fm in its ground state. 
In the absence of dissipation and fluctuations only symmetric fission
will be obtained with a rather narrow total kinetic energy (TKE)
distribution, weakly dependent on initial conditions.  The mean of the
TKE distribution in a hydrodynamic approach is significantly smaller
than the observed one, as the fissioning nucleus develops unexpectedly
long thin necks, reminiscent of the nuclear shapes obtained in the
liquid drop model with a large viscosity~\cite{Davies:1976}.
Similar longer necks develop if one were to increase significantly the magnitude 
of the pairing strength, when the dynamics of becomes 
very similar to the dynamics of perfect fluids~\cite{Bulgac:2018}. At the same time the width of 
the TKE distribution ($\approx 20$ MeV) is comparable to the observed one 
and to the numbers reported in Ref.~\cite{Tanimura:2017}.
Within that the nuclear shape develops a longer neck as the viscosity increases  
in the numerical implementation of the hydrodynamical approach of 
Ref.~\cite{Davies:1976}, a result at odds with our findings. 
\textcite{Davies:1976} restricted the nuclear shape to a parametrization using 5 DoF only, 
while in the present work we have included all shape DoF. These authors also 
expressed some doubts concerning the accuracy of the calculated 
inertia and viscosity tensors at large deformations. 
Shell-effects and pairing correlations can be accounted for by a 
variation of the macroscopic-microscopic 
formalism~\cite{BRACK:1972,Moller:2001}. From the known 
neutron and proton number densities one can construct the single-particle 
nucleon Hamiltonian, including spin-orbit and pairing interactions, and 
subsequently determine the 
corresponding energy density, 
which can then be used for the next time-step in Eqs.~\eqref{eq:td}. 
The full TDDFT description is likely a more efficient solution however. 

\begin{figure}
\includegraphics[clip,width=1\columnwidth]{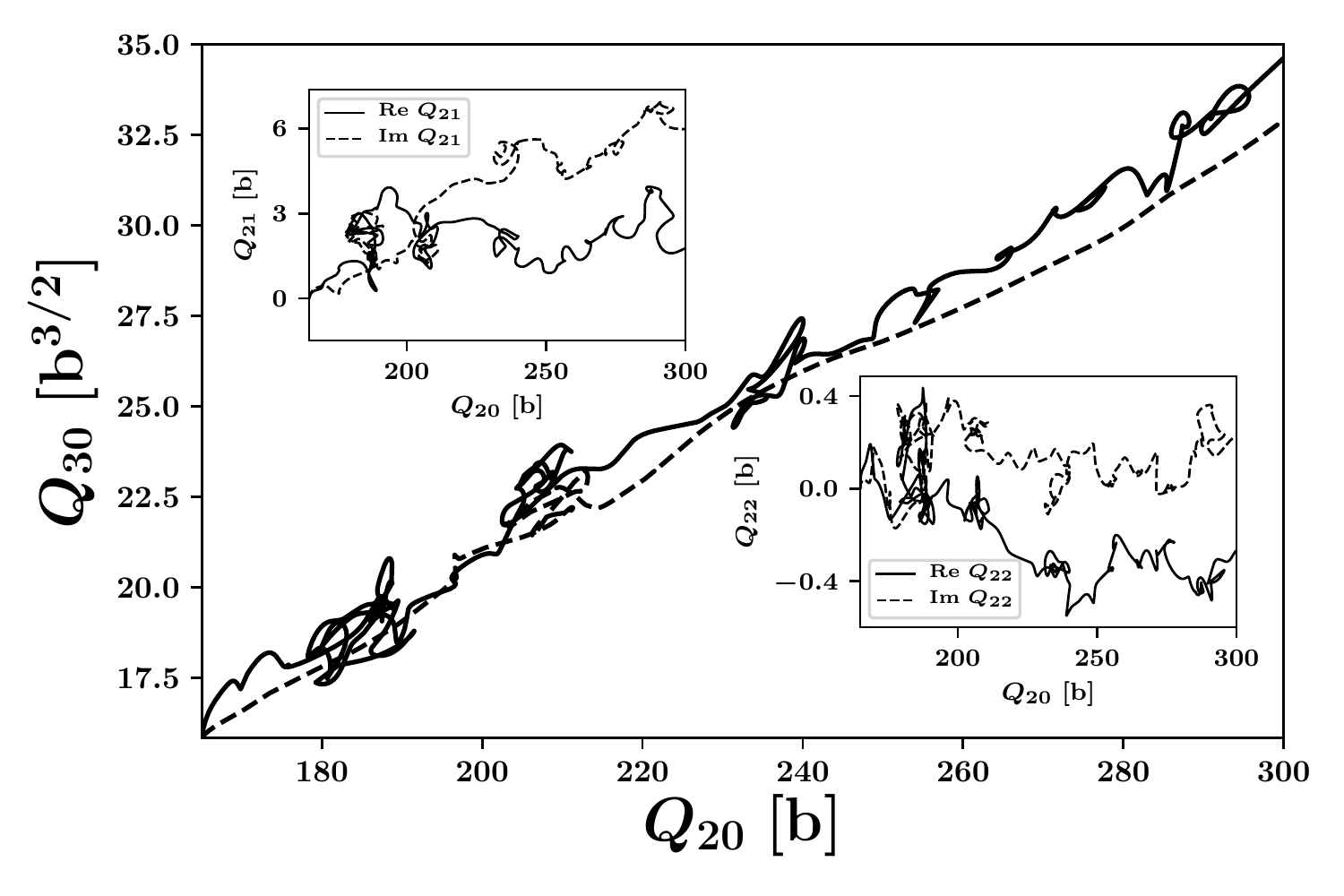}
\caption{\label{fig:tdslda}
In the main panel we show two typical full TDDFT trajectories for $^{240}$Pu  projected into the 
$ Q_{20}=\langle 2z^2-x^2-y^2\rangle  $  and  $Q_{30}=\langle (5z^2-3r^2)z\rangle$  plane obtained by evolving in time the TDDFT 
equations~\cite{Bulgac:2016,Bulgac:2018} without (dashed line) and
with (full line) dissipation and fluctuations included, using the NEDF SeaLL1~\cite{Shi:2018,Bulgac:2018}.  
In the insets we show the fluctuations of the moments $Q_{2m}= \langle z^{2-m}(x+iy)^m\rangle$, 
for $m=1,  2$, which vanish in the absence of fluctuations and otherwise break axial symmetry.
}
\end{figure}

For realistic calculations one has to resort to the full TDDFT
description~\cite{Bulgac:2016, Bulgac:2018}, using the evolution
equations \eqref{eq:fd}, with quasi-particle wave functions 
with spin and isospin DoF, and pairing correlations  accounted for.  
Such calculations require 
the use of  leadership computing facilities at a scale 
above that recently reported~\cite{Bulgac:2018}. If one were to resort to
a simpler approach, in which pairing correlations are described only at the BCS level, 
the calculational complexity is comparable to the a 
few hundred time-dependent Hartree-Fock trajectories needed
to perform ensemble averages, and such simulations are quite feasible~\cite{Tanimura:2017}.
In Fig.~\ref{fig:tdslda} we show a projection
onto the planes $\left ( Q_{20}, Q_{30}\right )$ and
$\left ( Q_{20},Q_{2m}\right )$ with $m=1,2$ of a typical full TDDFT
trajectory obtained without dissipation and fluctuations and including
dissipation and fluctuations using Eqs. \eqref{eq:fd}, in the case of
induced fission of $^{240}$Pu.  
Fluctuations breaking axial symmetry have never been considered in Langevin 
type of simulations and no informations was ever presented on their magnitude and importance.
To this end we integrate in time 442,368 complex coupled 
nonlinear stochastic partial differential equations on a 3D  
$24^2\times48$ spatial lattice with a lattice constant 1.25 fm, 
a time step 0.03 fm/c for 130,255 time-steps.
Only 6 DoF are illustrated in this
figure, even though all collective DoF were allowed to fluctuate. This simulation of a 
unitary quantum evolution including dissipation and fluctuations is exceeding by
$\approx {\cal O}(10^6)$  orders of magnitude any other simulation ever reported in 
literature in any physics field.\\

{\bf \emph{Conclusions}}\\

The main features of the present extension of the TDDFT formalism 
are: i) the present formalism is  quantum; ii) it includes all
collective DoF; iii) evolution is unitary in spite of including
explicitly dissipation; iv) all meanfield symmetries are broken
during the evolution, as expected for example in a full path-integral
description of the dynamics of an interacting many-fermion
system~\cite{Negele:1988,Bulgac:2010}, while in Langevin description
for example, axial symmetry was never broken; v) upon the inclusion of
fluctuations in TDDFT, the fission dynamics remains overdamped as
established in Ref.~\cite{Bulgac:2018}, collective kinetic energy (not
shown) remains as small as in their absence, and trajectories become
more convoluted and
longer in length and time and more random in the collective space; vi)
the meanfield adjusts naturally to the changes in the nuclear
shape; vii) without fluctuations one obtains only a lower limit of
fission times~\cite{Bulgac:2016,Bulgac:2018}. 
The formalism 
described here is applicable to many other situations: dissipative heavy-ion collisions,
non-equilibrium phenomena in cold atom physics; dynamics of vortices 
in neutron star crust, quantum turbulence~\cite{Bulgac:2016x}.\\

{\it Acknowledgments.}
  We thank M.M. Forbes for discussions and suggestions on the draft of
  the manuscript, Y. Akamatsu for discussions, and K.J. Roche for 
  help in using GPUs judicially.  The work of AB and
  SJ was supported in part by US DOE Grant No.~DE-FG02-97ER-41014 and
  in part by NNSA cooperative agreement DE-NA0003841.  The work of
  I.S. was performed at Los Alamos National Laboratory, under the
  auspices of the National Nuclear Security Administration of the
  U.S. Department of Energy.  Full TDDFT calculations have been
  performed by SJ at the OLCF Titan and Piz Daint and for generating
  initial configurations for direct input into the TDDFT code at OLCF
  Titan and NERSC Edison. This research used resources of the Oak
  Ridge Leadership Computing Facility, which is a U.S. DOE Office of
  Science User Facility supported under Contract No. DE-
  AC05-00OR22725 and of the National Energy Research Scientific
  computing Center, which is supported by the Office of Science of the
  U.S. Department of Energy under Contract No. DE-AC02-05CH11231.  We
  acknowledge PRACE for awarding us access to resource Piz Daint based
  at the Swiss National Supercomputing Centre (CSCS), decision
  No. 2016153479. This work is supported by "High Performance Computing 
  Infrastructure" in Japan, Project ID: hp180048. Hydrodynamic simulations 
  were carried out on the Tsubame 3.0 supercomputer at Tokyo Institute of Technology.
  This research used resources provided by the Los Alamos National Laboratory 
  Institutional Computing Program, which is supported by the U.S. Department of 
  Energy National Nuclear Security Administration under Contract No. 
  DE-AC52-06NA25396.


\providecommand{\selectlanguage}[1]{}
\renewcommand{\selectlanguage}[1]{}

\bibliography{local_fission}

\end{document}